# A New Method for Patterning Azopolymer Thin Film Surfaces

Sh. Golghasemi Sorkhabi[a,b], R. Barille[b], S. Ahmadi-Kandjani[a], S. Zielinska[c] and E, Ortyl[c]

[a.] *Research Institute for Applied Physics and Astronomy (RIAPA), University of Tabriz, Tabriz, Iran (Email: Shahla.golghasemi@gmail.com)*
[b.] *University of Angers/UMR CNRS 6200, MOLTECH-Anjou, 49045 Angers, France*
[c.] *Wroclaw University of Technology, Faculty of Chemistry, Department of Polymer Engineering and Technology, 50-370 Wroclaw, Poland*

Corresponding Authors
Shahla Golghasemi Sorkhabi
Tel: +98-413-3393019
E-mail: Shahla.golghasemi@gmail.com
Sh.Golghasemi@tabrizu.ac.ir

**Abstract**: We present a simple bottom-up approach via an incoherent unpolarized illumination and the choice of a solvent-droplet-induced-dewetting method to photoinducenano doughnuts on the surface of azopolymer thin films. We demonstrate that doughnut-shaped nanostructures can be formed and tailored with a wide range of typical sizes, thus providing a rich field of applications using surface photo-patterning. Furthermore, due to the presence of highly photoactive azobenzene derivative in the material, illumination of these nanostructures by a polarized laser light shows the possibility of a further growth and reshaping opening the way for fundamental studies of size-dependent scaling laws of optical properties and possible fabrication of nano-reactor or nano-trap patterns.

Keywords: Surface patterning, Azopolymer, solvent-droplet-induced-Dewetting, Photo-patterning, Nano-doughnut

# 1. Introduction

As a substantial base for the development of future technologies we are currently witnessing an explosion of novel ideas and strategies in nano-science with fusion of bottom-up and top-down strategies. Manipulation, conception and examination of nanostructured objects and devices in precise, sensitive and specific manners are some of the pillars of construction of new technologies for domains of photonics or nano-medecine[1]. In this regard, due to their potentials in the emerging field of controlled nanostructure formation, thin films of polymers containing azobenzenechromophores have generated significant interest with the development of nanofabrication and characterization techniques [2]. The geometrical configuration of the azo bond in azobenzene based compounds can be changed reversibly from *trans* to *cis* by irradiation with light intensity and/or polarization gradient. The photo-induced reversible *trans-cis-trans*isomerization of azobenzenes lead to spontaneous large scale macroscopic motion of polymer material leading to surface deformation of material below $T_g$[3]. This light induced movement has been recognized as a useful tool to enforce reversible changes in a variety of molecular systems; a possible technology for a broad range of fundamental and applied researches [4]. The mass transport not only provides a unique opportunity for nanostructure formation but also due to the unique photophysical behavior of these materials, azopolymers nanostructures can be optimized and reshaped via controlled light field, to gain the desired behavior [5].

In order to procure and explore novel opportunities for developing new applications in material science, nano-photonics and nano-biotechnology, advancement toward smaller features is of paramount importance in nano-fabrication [6].

In this regard, over the past decade, a variety of both top-down and bottom-up fabrication approaches such as: direct-writing, nano-imprinting and self-assembly, have been used to fabricate a range of well-defined nanostructured materials with desirable physical and chemical attributes. Among these, the bottom-up self-assembly process offers the most realistic solution toward the fabrication of next-generation functional materials and devices [7]. As a potential self-assembly technique, dewetting is regarded as a suitable method for micro and nanoscale fabrication as self-organization during dewetting leads to the fabrication of a nearly equal sized collection of holes and droplets [8- 9].

Several approaches have been developed for fabricating nanostructured arrays with organic molecules by controlling the dewetting[10-11]. Dewetting of polymer thin films have been explored experimentally and theoretically, although dewetting of azopolymer has not been suggested in the literature. Azopolymer nanostructures are produced through irradiation with a single laser or a laser pattern. However, examples of quite rare utilization of incoherent white-light for thin film regular photo-pattering and reshaping of azopolymer nanostructures has been recently demonstrated [12].

In this work, we experimentally show a simple bottom-up approach to produce doughnut shaped nanostructures on the surface of azopolymer thin films by the choice of a solvent-droplet-induced-dewetting method and an incoherent unpolarized light illumination. Also, due to the presence of highly photoactive azobenzene derivative in the material, illumination of these nanostructures by a polarized laser light shows the possibility of a further growth and reshaping of the structures.

## 2. Experimental

Azopolymer thin films are made from a highly photoactive azobenzene derivative containing heterocyclic sulfonamide moieties (IZO-3). The details of synthesis of the used copolymer based on 2- {2-[{4-[(E)-(4-{[(2,6-dimethyl-pyrimidin-4yl)amino]sulfonyl}phenyl)diazenyl]phenyl}(methyl) amino]ethoxy} ethyl 2-methylacrylate_ are reported elsewhere.The chemical structure of this copolymer is shown in Fig.1[13]. Thin films were prepared by dissolving azopolymer in THF (50 mg in 1 ml of THF) and spin-coated on a pre-cleaned glass substrate. Prepared films were let in oven overnight at 80ºC to remove any residual solvent and also for obtaining films whose surface morphology appears featureless or is not dominated by pinholes and other surface defects. The film thickness was determined by a Dektak Profilometer and was around 550 - 600nm. Mean molecular weight, $W_M$, of polymers has been determined by Gel permeation chromatography(GPC) using Waters 917 columns, RIDK-102 detector and APEX ver. 3.1 recorder and is between 14000 and 19000 g/mol. A mobile phase was γ- butyrolactone and molecular weight refers to polystyrene standards($\overline{M}_w$ =16500 g/mol, $\overline{M}_n$ = 11750 g/mol).

A Mettler Toledo DSC has been used for glass transition temperature (Tg) determination of polymers with scanning of 20K/min.The glass transition temperature ($T_g$) was 71˚C(344.5K).

Considering the importance of surface modification and the numerous applications of thin azopolymer thin films, this azopolymer and its relating compounds have been the subjects of studies for investigating the surface deformation through laser illumination [14-16]. However,the method chosen to create the initial patterns on the surface of azopolymer filmhere is solvent-droplet-induced dewetting. A 0.3 μl droplet of solvent was dropped on the thin film surface.Upon contact the droplet spreads and covers an area of approximately 50mm$^2$on the surface. An incoherent white light from a xenon lamp (Hamamtsu, C2177-01) was used to initiate the photoinduced mass transport in the film. To illuminate the sample by a polarized light, a horizontally linearly polarizedbeam from a DPSS laseroperating at a wavelength of λ = 473 nm was used. In the next steps a quarter/half wave plate was used to control and change the polarization and direction of the polarization of the laser beam. The thin film topography was studied with an atomic-force microscope (AFM, Veeco Instruments Inc) in the contact mode.

## 3. Result and discussion

We tested several common solvents as hexane, cyclohexane, toluene, chloroform, dichloromethane, dimethyl sulfoxide (DMSO), 1-4 dioxane and aceton. Finally n-Heptan was chosen for the optimal pattern on the surface in term of available structures. In a first experiment, a 0.3 μl droplet of the selected solvent (n-Heptan) was dropped off on the surface of the azopolymer thin film. A droplet of this volume covers an area of approximately 50mm$^2$on the surface.n-Heptan has a medium boiling point of 98˚C and an evaporation rate of 2.80 (Butyl acetat:1). Owing to the low solubility of azopolymer in the solvent (n-Heptan), dewetted films appear completely flat and undisturbed to the naked eye and does not exhibit any changes in the absorption spectra(Fig.2). The glass transition temperature of the polymer decreases below the room temperature due to absorption of the solvent into the polymer matrix increasing the free volume of the polymer and resulting in an enhancement of the polymer molecule mobility. These molecules are now able to reorganize freely which in turn leads to the rupture and dewetting of the thin film (hole formation) [17]. A small quantity of solvent entering in the azopolymer thin film is not sufficient to induce enough chain mobility. The glassy state is maintained and because the polymer is neither removed nor evaporated, the dissolved polymer can only be redistributed. The hole formation is due to migration of the polymer from the center of the hole to the perimeter during solvent evaporation[18].

The location of holes appears randomly but can be controlled with initial changes of the film topography [19] or by dewetting on already prepared patterned surfaces as we demonstrate later.

The mean length scale of the period between nearest nano-holes created by solvent-droplet-induced dewetting on the surface of azopolymer thin film was calculatedvia statistical processing of obtained AFM image (Fig.3). We found an average value $\lambda_f$ of 2.5 ± 0.1 μm which is in good agreement with the theoretical results of 2.17 μm, given by [11]:

$$\lambda_f = \left(\frac{M_e \gamma_s}{|S|}\right)^{1/3} \frac{h_0^{7/6}}{M^{1/2}} \tag{1}$$

With M is the PMMA polymer weight, $h_o$ the film thickness (600nm), $M_e$ the molecular weight between entanglements (7000 Da), $|S|$ the spreading coefficient (70 MPa) and $\gamma_s$ the PMMA surface tension (42 mJ/m$^2$). We consider in the calculation that PMMA has the largest matrix of material so the material was considered as pure PMMA in a simple model.

Multiple *trans−cis−trans* photo-isomerization cycles of chromophores in azopolymer induce a mechanical stress and deformations [20]. Studies have recently showed that a well-organized pattern on the surface of an azopolymer film can be formed by the use of an incoherent light source. Incoherent light can indifferently propel the photo-patterning process, in comparison with a laser illumination. Using this technique, it is possible to modify the holes and their surrounding areas formed on the surface evenly and to induce a polymer growth due to a mass transport [21].

The incoherent white light, from a xenon lamp with the effective power of 180 mW/cm², was used as an initiating source for a photo-induced mass transport in the film. To prove the ability of photo-induced nanostructuring, a detailed AFM scan of the film, before and after illumination, was done. Numerous randomly placed doughnut shaped nanostructures were clearly distinguished on the film. AFM images show exclusively doughnut shaped nanostructures with diameters varying from 150 nm to 500 nm. Using this typical film topography one can benefit from this variation of diameters in any possible applications requiring a simultaneous utilization of similar nano-objects with different sizes and depths (Fig.3).

The average diameter of the doughnuts is determined from peak-to-peak distances. A detailed analysis (using software WSxM) shows that the average diameter of the photoinduced doughnuts is typically 350 ± 50 nm (Fig.3).

The distribution of holes depths on the surface after dewetting seems almost constant (40 ± 10 nm). The illumination time for the sample was 30 min. During this time the force (F) needed to exert an elastic deformation on the nano-object is derived from the Hertz theory, which considers the contact deformation of elastic spheres under normal loads in absence of adhesion and fraction:

$$F = \frac{4}{3} G D_0^{0.5} \Delta D^{1.5} \qquad (2)$$

Where, G is the elastic modulus of PMMA (1.8 - 3.1), $\Delta D = D_0 - D$, the deformation where $D_0$ and D are the initial and final radius of the structure, respectively. Considering the same 1 GPa low bound, an elastic force of 22 µN was acting on the holes to produce a nano-doughnut. The volume of the doughnuts after illumination grows to reach as average amount of 0.134 µm³, which results from a growth rate of 0.0032 µm³, considering the 30 minutes of illumination time to reach the saturation point.

The average distance between the nearest nano-doughnuts is of 2.5 µm and is greater than their own diameters. This feature of the film makes it possible to work with nano-doughnuts in applications as isolated objects. Beside the wide range of possible geometrical sizes during fabrication, azopolymernano-doughnuts exhibit a property of flexible reshaping under illumination. To explore this property, the initial surface modification with holes acts as a mask for the photofluidization of these nanostructures. The structures were exposed to a horizontally linearly polarizedbeam from a DPSS laseroperating at λ = 473 nm, a wavelength close to its absorption maximum. The optical manipulation of polymer structures was done by varying the polarization and irradiation time, allowing an exceptional control of structural features.

Change and control of laser's polarization and direction of polarization was done using a quarter/half-wave plate. Samples were set perpendicular to the incident laser and illuminated by different polarizations (vertical, horizontal and circular polarizations) for a defined amount of time. It is seen that the initial symmetry ofthenano-doughnut changes with illumination, depending on the direction of polarization and duration of irradiation.

When the initial symmetrical doughnuts are irradiated with a linearly polarized light (horizontal polarization), the symmetry of these structures is modified. Nano-doughnuts shapes are changed into two semi-rings with a central hole (Fig.4). Under illumination the *cis-trans*photo-isomerization cycle of azopolymers gives rise to a light induced mass transport parallel to the light polarization. The molecules under irradiation are aligned perpendicularly to the incident polarization. This effect unravels the rim around the central hole. The variation of the two axial components (x or y) is not similar. The molecules along the x-axis move along the polarization direction, while the two semi-rings perpendicular to the polarization's direction build up. As a result, two poles of the doughnuts rim are voided and the two semi circles grew.

Such structures resemble a nano-cavity made with two semi-circular mirrors. An interesting application could be the possibility of creating cavities with nano-particles embedded in the center or be used to aggregate nano-particles along the rim [22].

Interestingly, if we continue the photo-reconfiguration by illuminating the holes with a linearly polarized light, the changes also occur along the vertical direction as well as the lateral direction. So, a further irradiation with a vertically polarized light (vertical polarization) after an irradiation with a laterally polarized light (horizontal polarization) onto doughnuts leads to a transformation of a circular hole into a square hole with an average diameter of 450 – 500 nm, as shown in the fig 5.a-b.

When the illuminating time is increased (up to 10 min), changes in the shapes are even more severe. The rectangular shape of the initial nano-doughnut reaches an average length of 850 nm and a width of 450 nm (fig.6). These structures are current subject of interest in the field of surface plasmon behavior with rectangular hole arrays [23].

A further increasing of the irradiation time results in surface relief grating (SRG) formation and the formed SRG overcomes the nanostructures on the surface.

Moreover, changing the polarization of the incident light to circular polarization,via a quarter wave plate, leads to growth nano-doughnuts. The average height of nano-doughnuts are about 15 nm, but measurements on structures created via circular polarization of light shows that the typical depth of these structures range between 40 – 110 nm. The measured heights for rims of these objects above the film surface vary from 15 – 25 nm. All the informations mentioned above lead to the classification of these structures as nano-well (fig.7.a-b).

Detailed analysis show that, not only the structures change their shapes during illumination, but also due the mass transport of the azopolymer a change in diameter and depth is also observed. The average diameter of the nanostructures varies in a range from 215 nm to 320 nm. Furthermore, the average depth of the initial structures formed after white light illumination grows from 15 nm to an average amount of 40 nm.

The study reveals a facile way to produce easily reproducible and controllable surface structures. Moreover, the great possibility of using light to dynamically photo-induce a mechanical change of the azopolymer structures presents a significant advantage as functional materials.

Changing the irradiation wavelength in the azopolymer absorption band or far from it can switch the behavior of doughnut-shaped nanostructures and leads to a flexibility of reshaping initial surface modifications to stable transparent nano-objects, thus covering a wide range of opportunity of applications.Such nanostructures are of great interest in application fields of sensors [24], nano-plasmonics[25-26], plasmonic solar cells [27], nanoparticle trapping [28], photochemistry [29], nano-rectors etc. Also, Smart nano-doughnuts with controlled volumes can be exploited as chemical reactors for photo-catalytic and enzyme reaction. Conveniently, all the surface modifications can be reversibly changed to come back to the initial state by heating the sample above the glass transition temperature.

As mentioned, dewetting provides a simple, powerful technique for patterning nanoparticles, small molecules, and polymers. Due to the complexity of the dewetting process it seems impossible to control well-defined hole/droplet formation. However, by using a pre-patterned substrate, we are able to direct ordered hole formation during the dewetting process.

Unlike the previous research on controlled dewetting, in which the substrates were patterned through conventional lithography, microcontact printing (μCP) and vapor deposition methods [30], azopolymer films can be patterned readily via an optical patterning.

In this stage, the azopolymer films were illuminated through a double beam exposure experiment. The sinusoidal light interference pattern at the sample surface leads to a sinusoidal surface patterning, i.e., a surface relief grating (SRG). We applied the solvent-droplet-induced-dewetting method to this surface expecting a modification of the surface in the place where the solvent is confined. The surface is then illuminated with a white light to produce a change of the initialsurface pattern. The result is visible in the figure 8.a where the nano-doughnuts are arranged on the edge of the stripe at regular distances decorating the grating with sub-structures. We point out that the sizes of the nanostructures are smaller than the previously obtained nanostructures on a flat surface, due to the limited space for growth and confinement of the solvent. We continued in this direction, based on results demonstrating that azo compounds have proven to allow the inscription of multiple superimposed relief gratings providing more complex structures (Fig.8.b). These gratings were considered as the template for a further dewetting process. The solvent droplet follows the pattern of the film in the regular nano-cavities. On these regularly patterned SRGs, we observed a spontaneous alignment of holes at the center of the trenches (Fig.8.c). Applying solvent on the two superimposed 1D gratings leads to the formation of a 2D array of ordered holes transformed in nano-doughnuts on the surface of the film.

## 4. Conclusion

We have demonstrated a route to fabricate the ordered nano-patterns from dewetting. These results present an exciting opportunity for manipulating structures and properties of an azopolymer thin film on nanometer scale. In this work, we experimentally presented a facile bottom-up method for fabrication of doughnut shaped nanostructures. The technique chosen for this purpose is dewetting of azopolymer thin film via a solvent-droplet induced method. The produced nanostructures can be classified as nano-doughnuts, consisting of a hole and a surrounding ring. Such structures can be used in fields such as nano-cavities, nano-rectors and sensors. Furthermore, as a result of the unique property of azopolymer thin films leading to massive photoinduced macroscopic motions of polymer chains, nanostructures of azopolymers can be optimized and reshaped via tailored light fields to obtain different modified structures. The exposition of the above films to a coherent polarized beam with an appropriate wavelength results in controlled growth and modification of structures. The next step is the use of the unique property of self-patterning of azopolymer films with different solvents in different conditions of surface ordering and temperature in order to produce desired regular patterns with a controlled dewetting process. This will help to optimize the control of structures to the required geometric forms.

**Figure captions:**

**Fig. 1.** Chemical structure of azopolymer

**Fig. 2.** Absorption spectra of the Azopolymer thin films before and after dewetting

**Fig. 3.** A typical topography obtained by AFM and a typical corresponding height cross-section of one nanostructure.

**Fig. 4.** Topographical images of the azopolymer film after the laser illumination (horizontalpolarization) with the three-dimensional AFM image of nanostructuresformed on surface under laser radiation.

**Fig. 5.** (a) Topographical images of the azopolymer film after laser illumination with horizontal and vertical polarization, (b) Three-dimensional AFM image of nanostructures formed on surface of Azopolymer film under laser radiation.

**Fig. 6.** Topographical images of the azopolymer film after the laser illumination (horizontal polarization) up to 10mins with the three-dimensional AFM image of nanostructures formed on surface under laser radiation.

**Fig. 7.** (a) Topographical images of the azopolymer film after the laser illumination (circular polarization), (b) Three-dimensional AFM image of nanostructures formed on surface of Azopolymer film under laser radiation and corresponding height cross-section.

**Fig. 8.** (a) Dewetted structures on a 1D templated film, (b) AFM image of a 2D gratings, (c) Dewetted structures on a 2D templated film

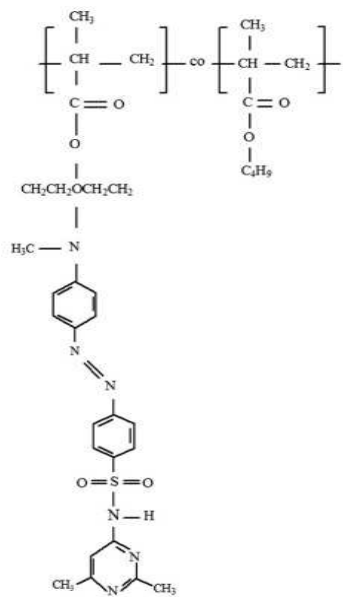

**Fig.1**. Chemical structure of azopolymer

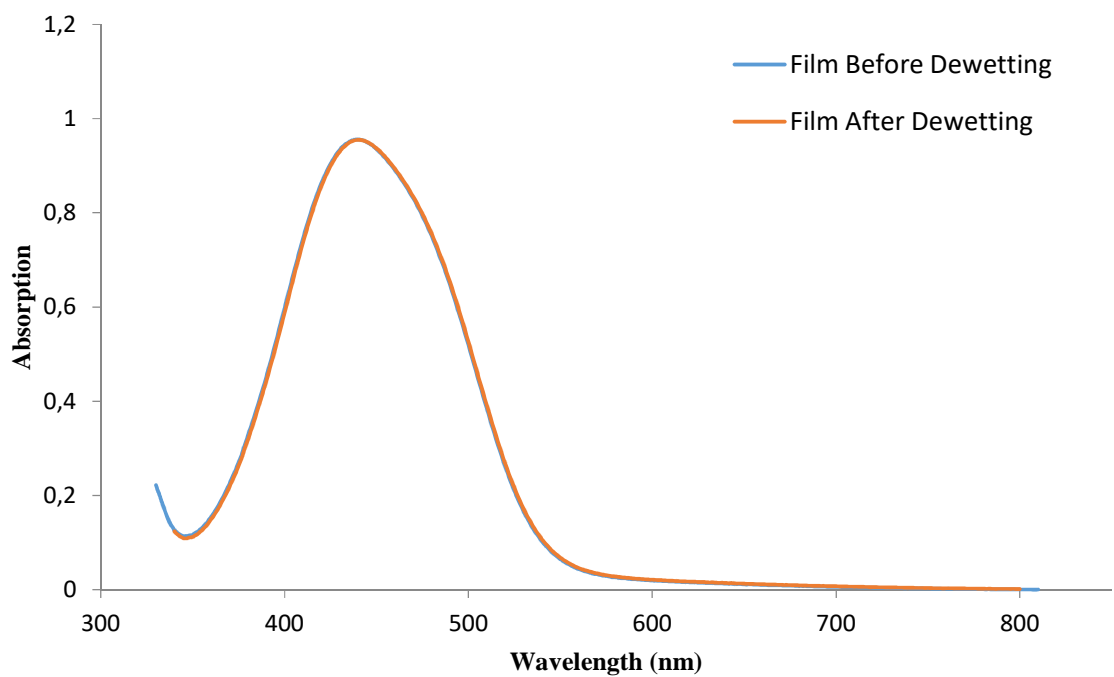

**Fig. 2.** Absorption spectra of the Azopolymer thin films before and after dewetting

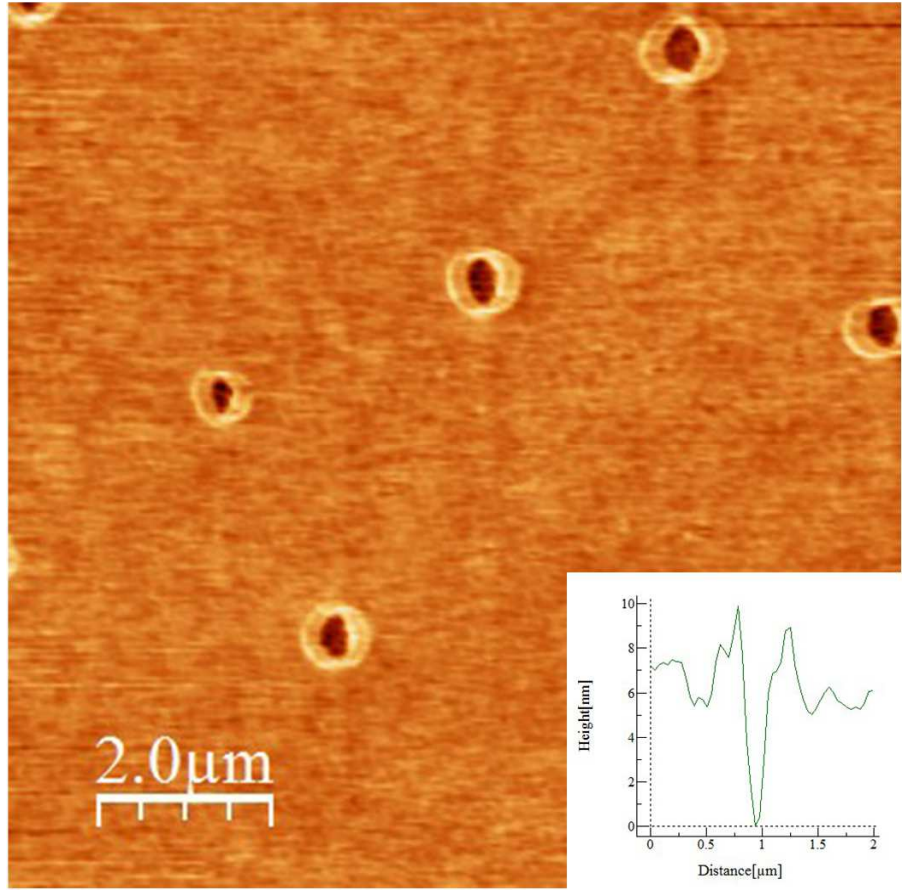

**Fig. 3.** A typical topography obtained by AFM and a typical corresponding height cross-section of one nanostructure.

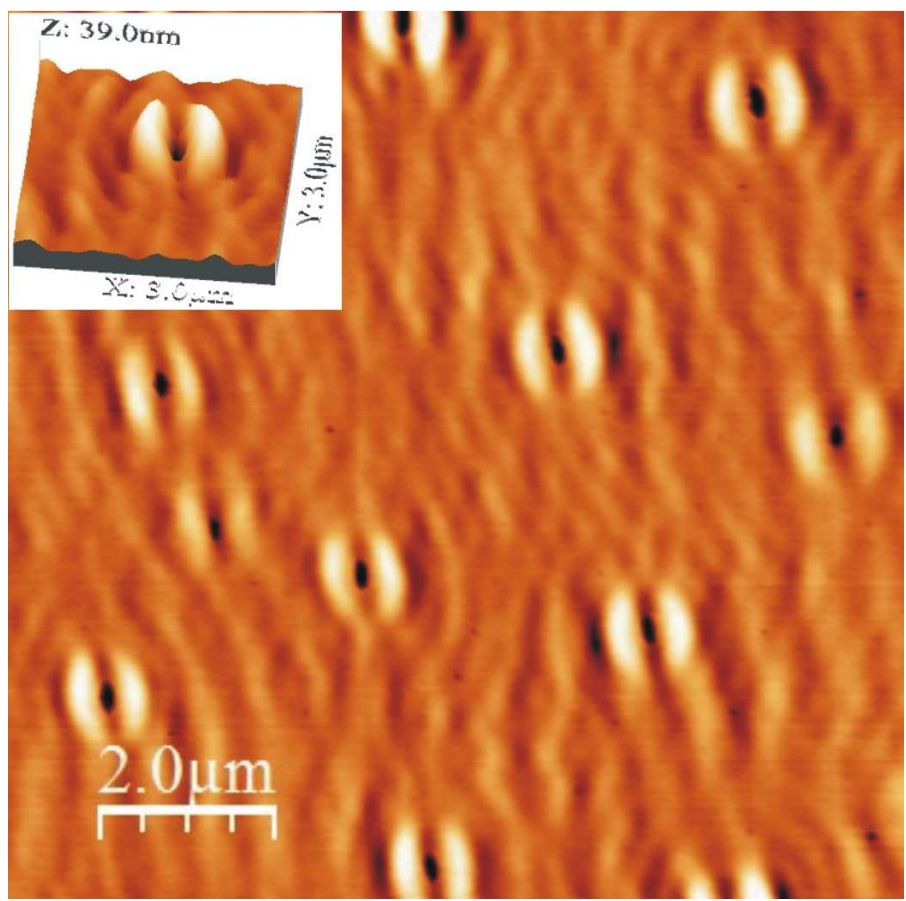

**Fig. 4.** Topographical images of the azopolymer film after the laser illumination (Horizontal polarization) with the three-dimensional AFM image of nanostructures formed on surface under laser radiation.

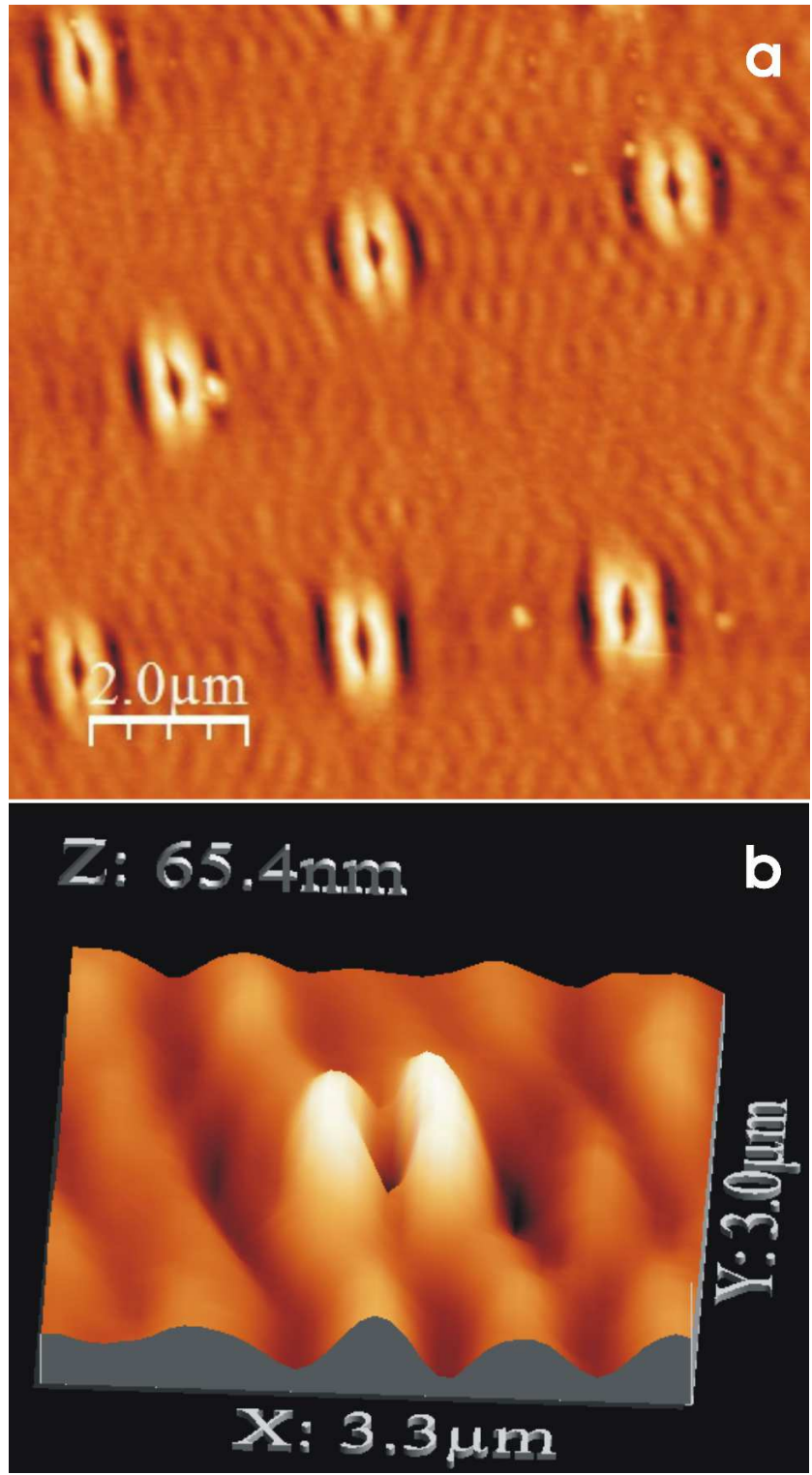

**Fig. 5.**(a) Topographical images of the azopolymer film after laser illumination with horizontal and vertical polarization, (b) Three-dimensional AFM image of nanostructures formed on surface of Azopolymer film under laser radiation.

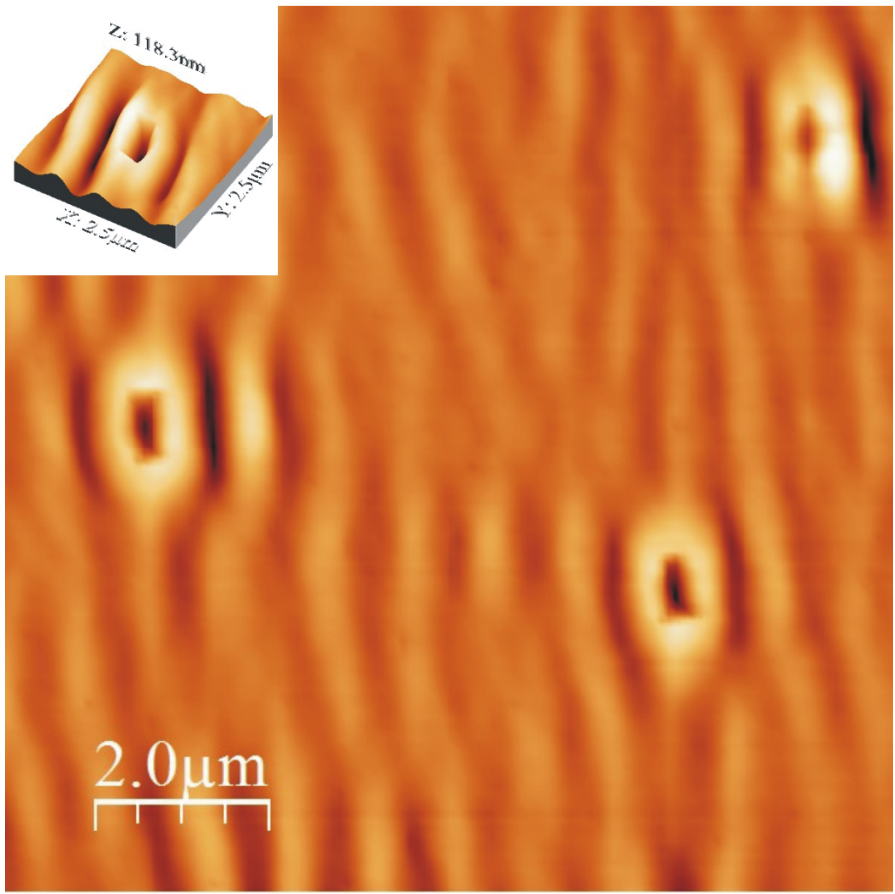

**Fig. 6.** Topographical images of the azopolymer film after the laser illumination (horizontal polarization) up to 10mins with the three-dimensional AFM image of nanostructures formed on surface under laser radiation.

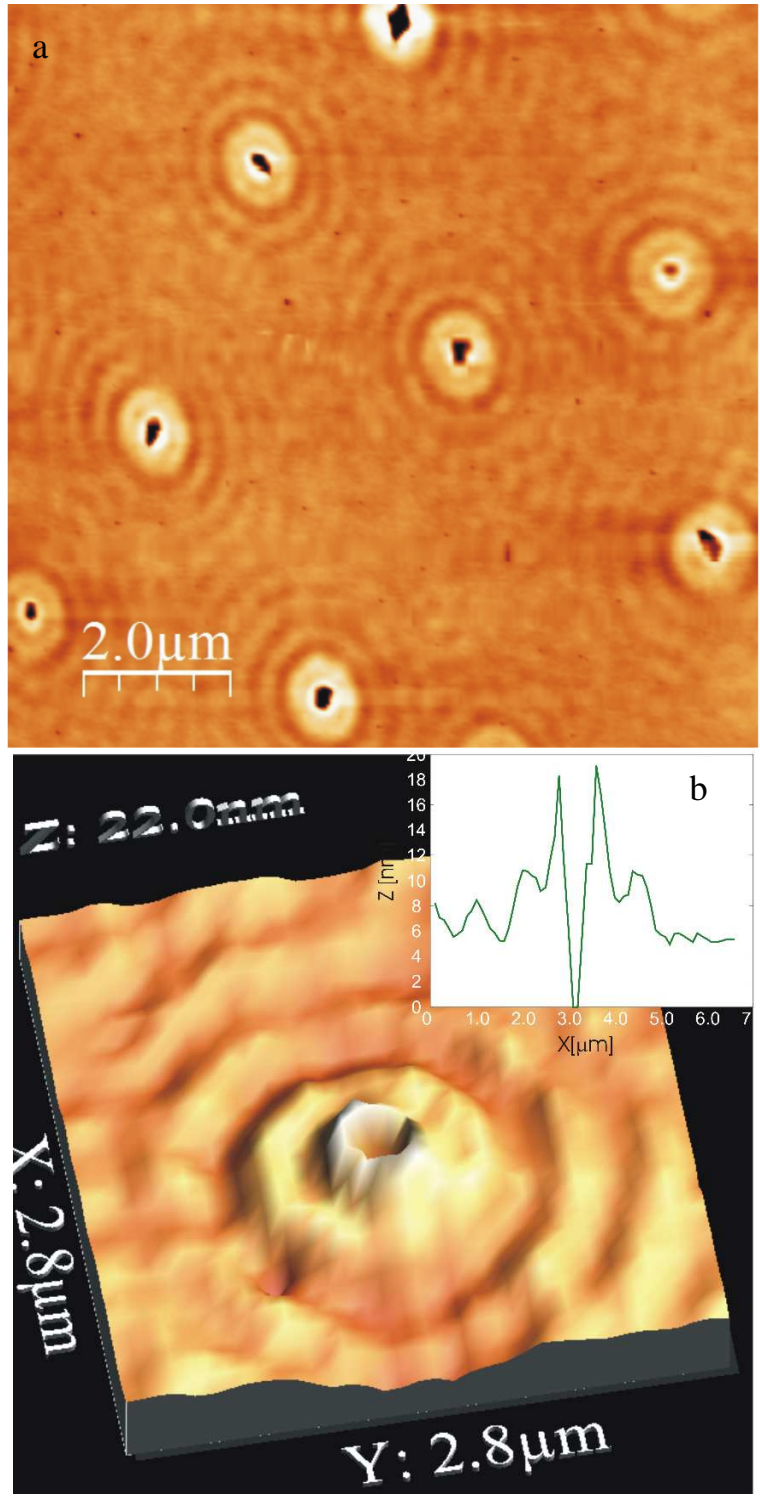

**Fig. 7.**(a) Topographical images of the azopolymer film after the laser illumination (circular polarization), (b) Three-dimensional AFM image of nanostructures formed on surface of Azopolymer film under laser radiation and corresponding height cross-section.

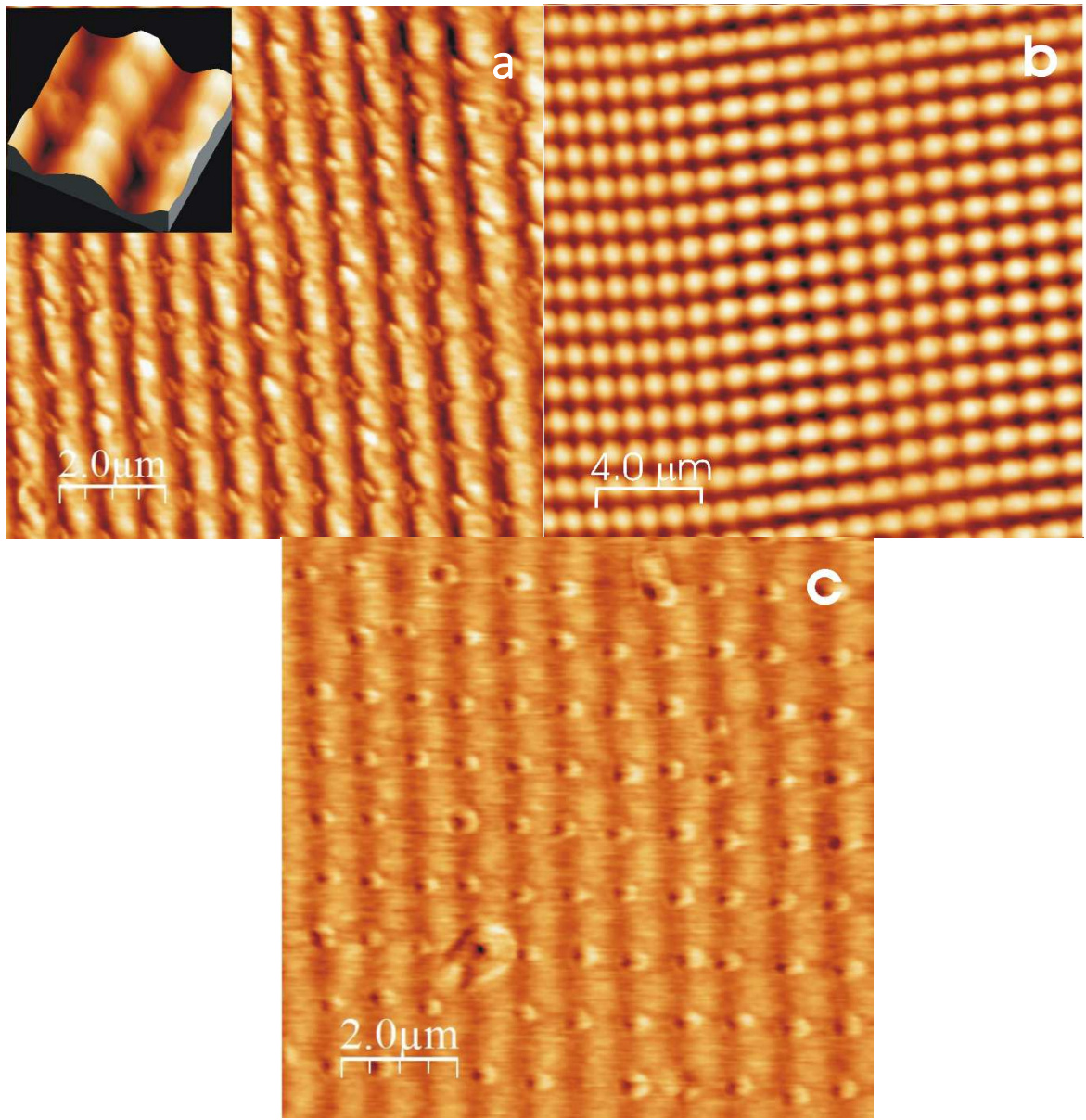

**Fig. 8.**(a) Dewetted structures on a 1D templated film, (b) AFM image of a 2D gratings, (c) Dewetted structures on a 2D templated film